\documentclass[aps,amssymb,amsmath,superscriptaddress,notitlepage,longbibliography,nofootinbib]{revtex4-1}

\usepackage{booktabs}
\usepackage{graphicx}
\usepackage{float}
\usepackage{color}
\usepackage{subcaption}

\newcommand{\ra}[1]{\renewcommand{\arraystretch}{#1}}

\AtBeginDocument{
\heavyrulewidth=.08em
\lightrulewidth=.05em
\cmidrulewidth=.03em
\belowrulesep=.65ex
\belowbottomsep=0pt
\aboverulesep=.4ex
\abovetopsep=0pt
\cmidrulesep=\doublerulesep
\cmidrulekern=.5em
\defaultaddspace=.5em
}

\begin{document}
\title{Critical role of electronic correlations in determining
crystal structure of transition metal compounds}

\author{Nicola Lanat\`a}
\email{lanata@magnet.fsu.edu}
\thanks{Contributed equally to this work}
\affiliation{Department of Physics and National High Magnetic Field Laboratory, Florida State University, Tallahassee, FL 32310}
\author{Tsung-Han Lee}
\email{tl596@physics.rutgers.edu}
\thanks{Contributed equally to this work}
\affiliation{Department of Physics and National High Magnetic Field Laboratory, Florida State University, Tallahassee, FL 32310}
\affiliation{Department of Physics and Astronomy and Center for Condensed
Matter Theory, Rutgers University, Piscataway, NJ 088548019}
\author{Yong-Xin Yao}
\email{ykent@iastate.edu}
\affiliation{Ames Laboratory-U.S. DOE and Department of Physics and Astronomy, Iowa State University, Ames, Iowa 50011}
\author{Vladan Stevanovi\'c}
\email{vstevano@mines.edu} 
\affiliation{Colorado School of Mines and National Renewable Energy Laboratory, Golden, Colorado 80401}
\author{Vladimir Dobrosavljevi\'c}
\email{vlad@magnet.fsu.edu}
\affiliation{Department of Physics and National High Magnetic Field Laboratory, Florida
State University, Tallahassee, FL 32310}



\raggedbottom
\maketitle
%
%
\thispagestyle{empty}

{\bf
The choice that a solid system "makes" when adopting a crystal structure (stable or metastable) is ultimately governed by the interactions between electrons forming chemical bonds. 
By analyzing 6 prototypical binary transition-metal compounds we demonstrate here that the orbitally-selective strong $d$-electron correlations influence dramatically the behavior of the energy as a function of the spatial arrangements of the atoms. Remarkably, we find that
the main qualitative features of this complex behavior 
can be traced back to simple electrostatics, i.e., to the fact that 
the strong $d$-electron correlations influence substantially the charge transfer mechanism, which, in turn, controls the electrostatic interactions. This result advances our understanding of the influence of strong correlations on the crystal structure, opens a new avenue for extending structure prediction methodologies to strongly correlated materials, and paves the way for predicting and studying metastability and polymorphism in these systems.
}

	
Predicting the ground-state structure of crystalline materials, initially thought to be an unsolvable problem, became an active area of 
research with the advent of efficient numerical implementation of computational total energy methods.
Various approaches to exploring
potential energy surface of solids (PES) from first principles
(ab-initio thermodynamics)
have been developed~\cite{catlow_NMAT:2008, oganov2010modern, evrenk2014prediction},
leading to exciting discoveries 
such as superconducting dense hydrogen sulfide~\cite{li_JCP:2014}, new and intriguing forms of matter at elevated pressures~\cite{na_Nature:2009}, new functional 
materials~\cite{hautier_CM:2011, peng_PRX:2015, anubhav_NRM:2016}, and other.
Beyond the ground-state structures, efforts in exploiting polymorphism and extending structure prediction and materials by design strategies to metastable systems
have also been pursued~\cite{botti_PRB:2012,huan_PRL:2013, Stevanovic_PRL:2016, jones_PRB:2017}.
Describing accurately the PES was proven 
essential also in the efforts to predict from first principles new
thermodynamically-stable
topological insulators~\cite{Zunger1,Zunger2} and Weyl-Kondo semimetals~\cite{Dirac-Kondo}.

The space of strongly correlated electron systems, on the other hand, represents a virtually untapped territory for finding new materials exhibiting
potentially ground-breaking physical properties.
However, exploring the PES of these materials 
(usually $d$- or $f$-electron systems) poses significant challenges.
This includes both the``choice'' of the ground state structure, and 
the energy ordering of different PES local minima.
The fact that strong electron correlations can influence
the ground-state crystal structure
has been demonstrated previously. 
For example, including correlations at an 
appropriate level of theory (Random Phase Approximation and Quantum Monte Carlo) was shown to be critical in reproducing the known rocksalt ground state structure of
MnO~\cite{Schroen_PRB:2010, Peng_PRB:2013, 
Schiller_PRB:2015}. Contrary to experiments, correlation-deficient approaches
such as the classic approximations to density functional theory 
(DFT)~\cite{HohenbergandKohn,KohnandSham}
would suggest zincblende or wurtzite structure, 
both featuring tetrahedrally coordinated atoms, to be lower in energy than the octahedrally coordinated rocksalt. 
However, the general
physical mechanism through which electron correlations influence 
relative energies of different crystal structures remains elusive, and
a systematic investigation of the role of correlations in determining the
thermodynamically-stable crystal structure of 
transition metal compounds is presently missing. 

In this paper we investigate the influence of strong electronic correlations present in $d$-transition metal compounds on the ``choice'' of the ground-state 
crystal structure of these systems. The main findings of our work are the following: (1) The strong electron correlations influence dramatically many important features 
of the PES in $d$-electron materials, such as the energy ordering of different polymorphs and the thermodynamically stable crystal structure.
(2) Available theories describing the strong electron correlations beyond a mean-field single-particle
picture, see, e.g., Refs.~\cite{LDA+U+DMFT,DMFT,dmft_book,LDA+U+DMFT,Fang,Ho,PRB_GA_SISSA,PRX_LDA_RISB,PRL_UO2_Lanata},
provide us with effective tools for simulation-based structure-prediction studies of $d$-electron materials. 
(3) The main physical mechanism underlying the interplay between strong electron correlations and crystal structure in transition metal compounds --- especially Mott insulators --- is 
charge transfer, which governs the electrostatic interactions, that, in turn,
play a decisive role in determining the ground-state structure.

\begin{figure}[H]
	\begin{center}
		\includegraphics[width=0.60\linewidth]{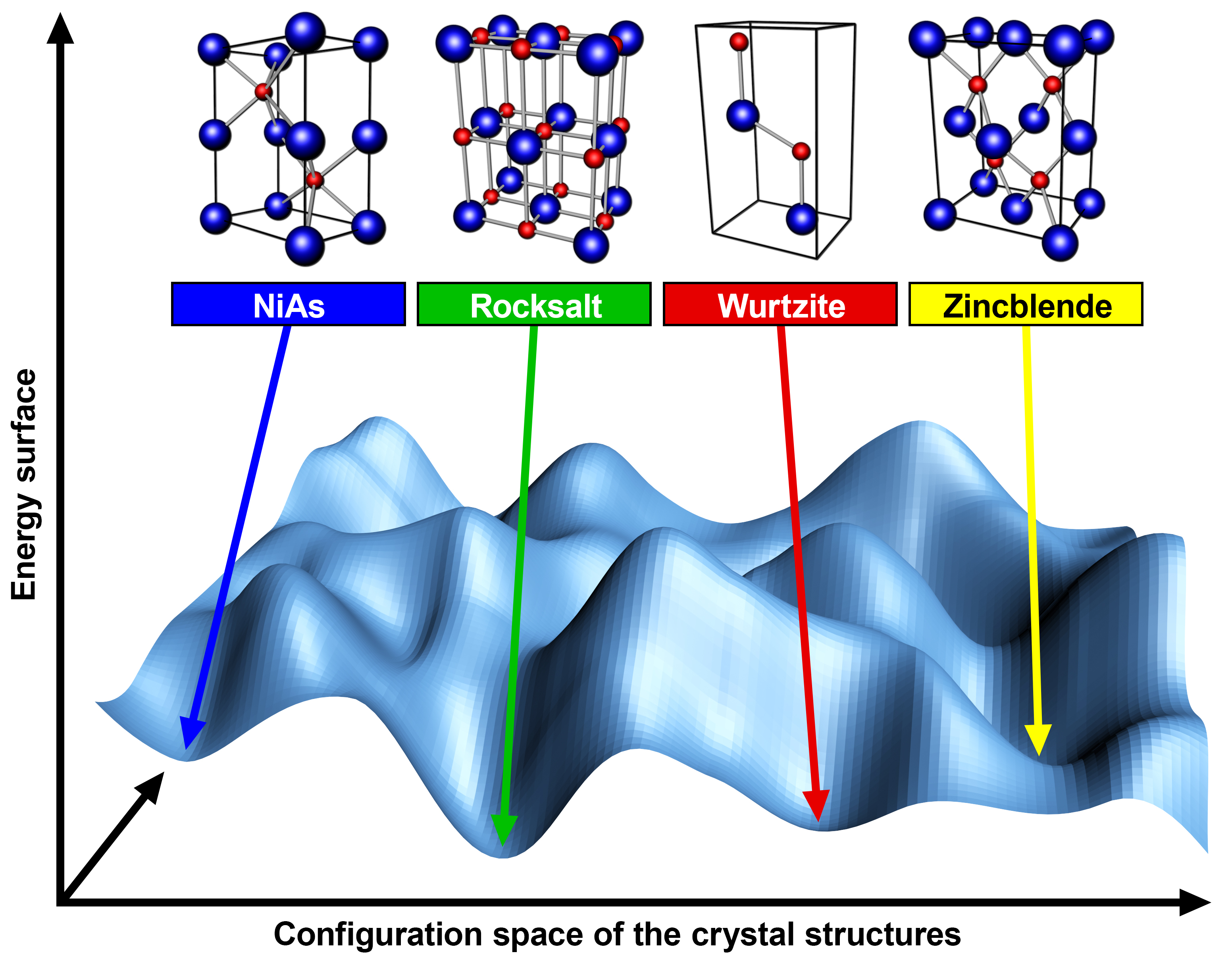}
		\caption{Representation of generic energy profile 
			as a function of the crystal configuration. Crystal structures considered in this work:
			NiAs-type, rocksalt, wurtzite, and zincblende.}
		\label{structures}
	\end{center}
\end{figure}

The claims above are demonstrated here by studying 6 transition metal binary oxides and chalcogenides (CrO,  MnO, FeO, CoO, CoS and CoSe) in 4 common crystal structure types
shown in Fig.\ref{structures} (rocksalt, NiAs-type, zincblende and wurtzite). This particular selection of crystal structures covers both the change in local 
coordination of the atoms and in the long range order, because it combines octahedral coordination with cubic symmetry (rocksalt),  octahedral coordination with hexagonal symmetry (NiAs-type), 
tetrahedral coordination with cubic symmetry (zincblende) and tetrahedral coordination with hexagonal symmetry (wurtzite).
Furthermore, the selected set includes the ground state structure for all studied systems. 
By employing a combination of local density approximation (LDA)~\cite{LDA}
and the rotationally-invariant slave-boson (RISB)
theory~\cite{rotationally-invariant_SB,RISB_Lechermann,PRL_UO2_Lanata}, which
includes electron correlations beyond 
the single particle picture \cite{Fang,Ho,rotationally-invariant_SB,Slave-bosons-materials,PRX_LDA_RISB,PRL_UO2_Lanata}, we were able to reproduce the experimentally known ground state structure of all compounds, as well as to uncover 
dominant physical mechanism by which strong correlations influence the energy ordering of different crystal  structures.

\section*{Results}
We have performed LDA, LDA+U~\cite{LDA+U} and LDA+RISB ground-state calculations of 
CrO,  MnO, FeO, CoO, CoS and CoSe in 4 different crystal structures
(rocksalt, NiAs-type, zincblende and wurtzite, see Fig.~\ref{structures}).
The LDA+RISB calculations have been performed assuming a Hund's
coupling constant strength $J=0.9\,eV$ and scanning different values of 
Hubbard interaction strength $U$. 
Since all of these transition-metal compounds  
are paramagnetic at room temperature, the LDA+RISB
simulations were performed assuming from the onset
paramagnetic solutions.

\begin{figure}[H]
	\begin{center}
		\includegraphics[width=0.99\linewidth]{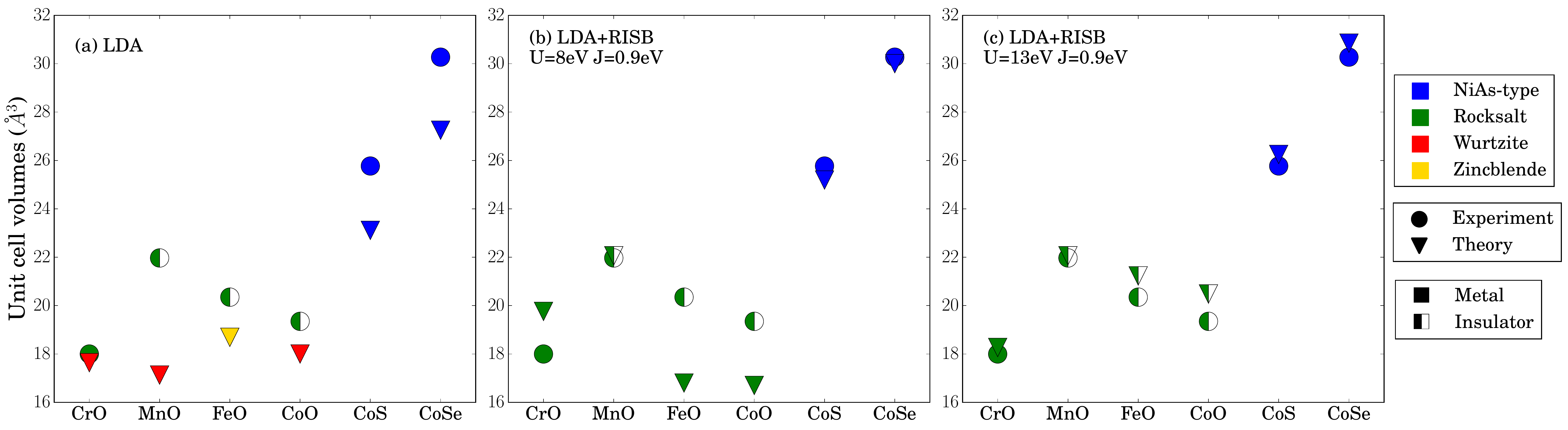}
		\caption{Theoretical (triangles) unit cell volumes and crystal structures in comparison with the experiments (circles).
			The colors, blue, green, red, and yellow, correspond to NiAs-type, rocksalt, wurtzite, and zincblende structures, respectively. The metallic and insulating solution are labeled by filled and half-filled symbols, respectively.
		}
		\label{volumes}
	\end{center}
\end{figure}

In Fig.~\ref{volumes} is provided a bird's eye view of 
the main properties of the materials considered,
inherent in their zero-temperature thermodynamically stable phase, i.e.: 
(1) the crystal configuration, (2) the equilibrium density,
and (3) whether the system is a metal or an insulator.
The theoretical results (triangles) are shown
in comparison with the experiments (circles).
The reported experimental data 
were obtained from Refs.~\cite{CrO,MnO,FeO,CoO,CoS,CoSe}.
%
The crystal structures are color coded by blue, green, red, yellow, for NiAs-type, rocksalt, wurtzite,
and zincblende, respectively.
The insulating phases are indicated by half-filled symbols,
while the metallic phases are indicated by fully filled symbols.

From the experimental data we observe that
all of the oxides considered favor the rocksalt structure,
while the thermodynamically stable lattice configuration of
CoS and CoSe is NiAs-type.
All of the materials have a metallic ground state except 
MnO, FeO and CoO, which are Mott insulators.

As mentioned in the introduction, 
LDA fails to reproduce the experimental crystal
structure (rocksalt) for all of the oxides.
The method is also unable to capture the fact that 
MnO, FeO and CoO are insulators, and the 
predicted equilibrium volumes are generally inaccurate.
Inclusion of spin polarization (ordering) at the level of LDA+U or (GGA+U)
straightens out the limitations of unpolarized LDA only in part.
In particular, as shown in the Supplemental Material, the spin polarized GGA+U
suggests different ground state structures of both CoS and CoSe depending on
the value of $U$, never reproducing the experimentally known NiAs structure as the lowest energy one. 
The lowest energy structure changes from the rocksalt derived ($U=0\,eV$) to zincblende ($U=6\,eV$) and wurtzite ($U=12\,eV$).
Furthermore calculated equilibrium volumes are very different from the experimental ones.
Hence, the experimentally observed change in the long-range order from the rocksalt
to NiAs-type structure, that preserves octahedral coordination of atoms and occurs as the anion is replaced from CoO to CoS and CoSe, is overall poorly described by 
spin polarized GGA+U (or LDA+U).
These results follow from an exhaustive enumeration of different spin configurations constructed on all symmetry inequivalent supercells with up
to four formula units (approximately 700 calculations for both CoS and CoSe). The spin polarized GGA+U calculations were performed following the approach 
described in Ref.~\cite{vladan_PRB:2012}, which employs PAW treatment of valence electrons~\cite{bloechl_PRB:1994} 
and the Perdew-Burke-Ernzerhof (PBE) functional form for the exchange-correlation functional within the generalized gradient approximation to DFT~\cite{perdew_PRL:1996},
as implemented in the VASP code \cite{kresse_CMS:1996}.
These results constitute unequivocal evidence of the fact that the strong electron correlations influence substantially the
behavior of both the electronic structure and the total energy of these materials.

Remarkably, the LDA+RISB theory provides results in very good
quantitative agreement with the experiments
for all 6 transition metal binary oxides and chalcogenides considered,
simultaneously.
The simulations are particularly accurate
for $U=13\,eV$.  In fact, for this value of the Hubbard interaction strength,
the method captures, at the same
time, all of the physical properties examined,
including the crystal structure and the insulating nature of 
MnO, FeO and CoO. Furthermore, 
the experimental equilibrium volumes of all materials
are reproduced within $4\%$ error.
We note that the method captures the correct crystal structure
of all materials also for $U=8\,eV$, although the overall accuracy
of the results is not as satisfactory as for $U=13\,eV$. 
In particular, the equilibrium volume of FeO and CoO is underestimated
for this value of $U$.
The reason why varying the value of $U$ influences considerably
the equilibrium volume for these 2 materials is that
the MIT of these 2 systems, which
occurs at $U\lesssim 13\,eV$, is first order. 
Consequently, the equilibrium volume evolves discontinuously
at the critical point \cite{CoO_LDA_DMFT_PRB,MnO_LDA_DMFT_nat}.
Note that the values of the MIT critical $U$ of FeO and CoO
reported above are likely overestimated
(this is a well-known systematic limitation of the RISB approximation).

These results indicate clearly that taking into
account the strong electron correlations beyond the single-particle picture,
e.g., utilizing the LDA+RISB approach, results in a remarkably
effective tool for structure-prediction work of $d$-electron materials.

\begin{figure}[H]
	\begin{center}
		\includegraphics[width=0.75\linewidth]{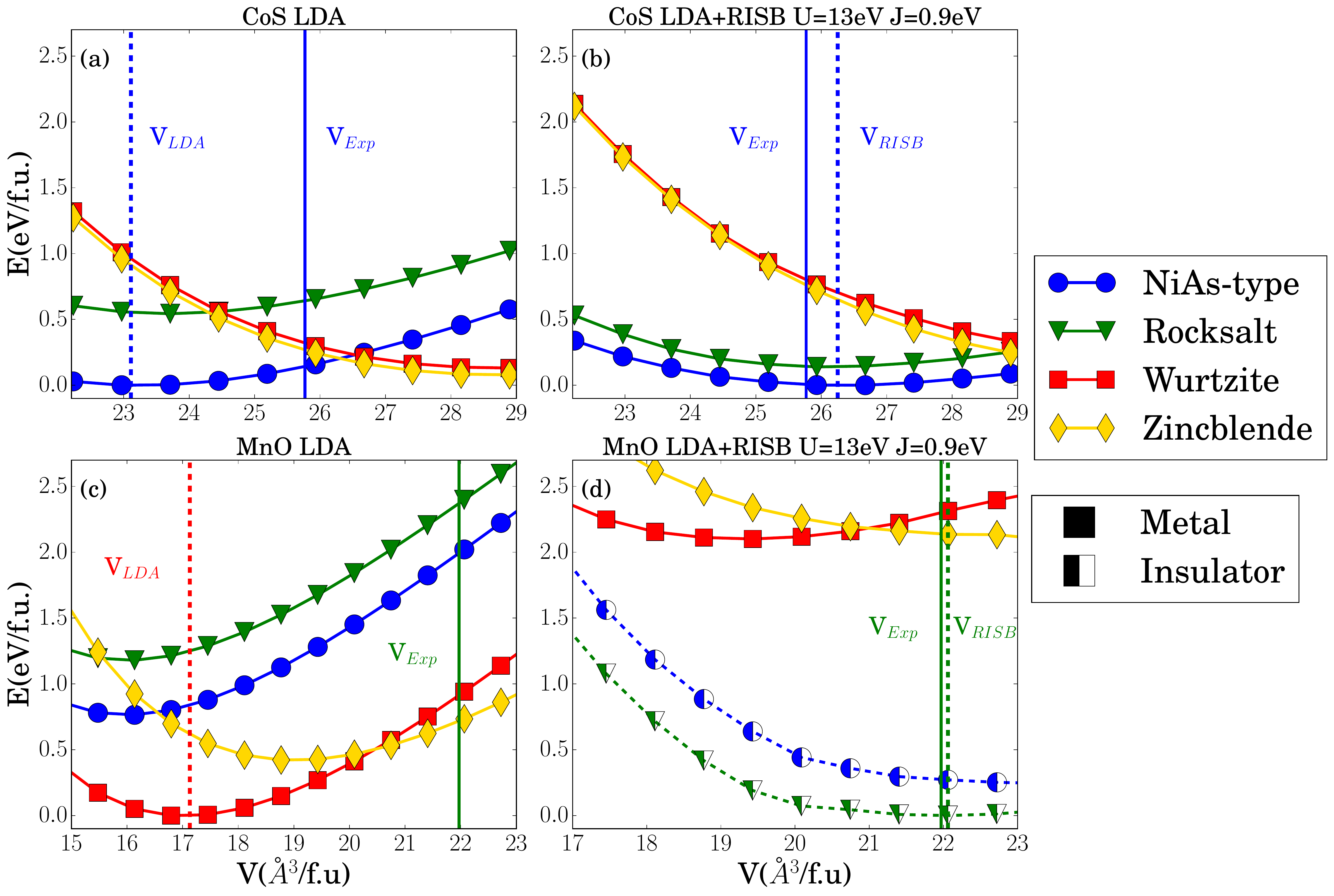}
		\caption{ 
			LDA and LDA+RISB behaviour of the energy as a function of 
			the volume for CoS (panels a,b) and MnO (panels c,d).
			The blue circle, green triangle, red square, and yellow diamond curves correspond to NiAs-type, rocksalt, wurtzite, and zincblende structures, respectively.
			The experimental equilibrium volumes for each compounds are marked 
			by vertical solid lines, while the theoretical equilibrium volumes are marked by in dashed lines.}
		\label{Fig3}
	\end{center}
\end{figure}

In order to 
investigate in further detail the influence on the crystal structure
of the strong $d$-electron 
correlations, and assess its relevance for the prediction of new polymorphs, 
it is also interesting to inspect
the energy profiles of the thermodynamically unstable
crystal configurations.  
For illustration purposes, the behavior of the theoretical
total energy as a function of the volume
is shown in Fig.~\ref{Fig3} only for CoS and MnO, 
both in LDA and in LDA+RISB. 
The analysis of the other materials is reported
in the Supplemental Material.

Interestingly, the RISB correction modifies dramatically 
the LDA energy order of the crystal structures,
both for CoS (which is a metal)
and MnO (which is a Mott insulator).
We note also that the LDA+RISB relative energies
between the different crystal structures
of both MnO and CoS, evaluated at their respective experimental equilibrium volumes,
are considerably larger with respect
to the N\`eel temperature of the
materials examined, that are all paramagnetic at room temperature.
In particular, the energy difference between the
octahedrally coordinated configurations
(Rocksalt and NiAs-type) and the
tetrahedrally coordinated configurations (Wurzite and Zincblende)
is of the order of $\sim 1\,eV$ for both of these materials.

The critical role of the strong electron correlations in determining
the crystal structure is well exemplified by the calculations
of MnO displayed in panel (d) of Fig.~\ref{Fig3}.
In fact, MnO is a Mott insulator in the octahedrally coordinated structures,
while it is a metal in the tetrahedrally coordinated structures.
As shown in the Supplemental Material, the same behavior is observed
in FeO and CoO, which (at their respective experimental equilibrium volumes)
are also Mott insulators in their thermodynamically
stable crystal configuration, while they are metals in the tetrahedrally
coordinated structures.

These results indicate clearly that computing the behavior of the
PES requires to take into account the subtle competing mechanisms underlying
the strongly-correlated regime around the Mott transition --- which 
can only be accomplished by
many-body techniques able to take into account the strong
electron correlations beyond a mean-field single-particle picture,
such as DMFT and the RISB.
In this respect, we note also
that the interplay between crystal structure and $d$-electron correlations 
in the materials examined here is considerably more complex 
with respect to $f$-electron systems such as elemental Pr and Pu,
where the RISB correction to the total energy 
was shown to be very similar for all phases~\cite{PRX_LDA_RISB}.

It is also interesting that turning on the $d$-electron Hubbard interaction 
has a qualitatively distinct effect on the relative energies
between different crystal structures
with respect to increasing the volume.
In fact, for all of the materials considered, 
the LDA Wurtzite and Zincblende energies
tend to become smaller than the Rocksalt and NiAs-type energies 
at larger volumes. 
Instead, turning on the interaction parameters $(U,J)$ has the opposite effect,
i.e., it favors energetically the Rocksalt and NiAs-type structures.
This consideration forces upon us the conclusion that
the observed pronounced physical effects 
of the strong electron correlations
on the behaviour of the total energy,
which are accounted for by the LDA+RISB theory,
are inherently "orbitally-selective"; in the sense that 
their influence on the crystal structure
can not be simply traced back into 
a uniform renormalization of the bandwidth,
as one might naively expect. 
As we are going to show, in fact, one of the main physical effects
at play is charge transfer, which is a  multi-orbital
phenomenon decisively influenced by the strong electron correlations.

In order to 
investigate the connection between Mott physics and the energy order of the crystal structures,
it is insightful to take a step back and analyze the energy-ordering problem from the point of view of a simple classic point-ion electrostatic (PIE) model.
The PIE model and the Madelung energy were successful 
in explaining the structure and the order-disorder phenomena in
spinels~\cite{Stev_PRL:2010,Stev_JACS:2011} as well as in non-isovalent perovskite alloys~\cite{PIE-Vanderbilt}.
Here we employ the PIE model and Ewald summation described in Ref.~\cite{Stev_PRL:2010}, and evaluate the 
Madelung energies at the experimental equilibrium volumes, using as inputs the theoretical local $d$-electron occupations $n_d$
reported in Table~\ref{n_d-table}. The point charges used to compute the Madelung energy are: 
(a) the transition metal sites are assumed to carry a positive charge
equal to the total number of valence electrons 
minus the value of $n_d$, and (b) the anion sites are assumed to
have a negative charge such as to make the system charge-neutral.

\begin{table}[H]\centering
\ra{1.3}
\caption{LDA and LDA+RISB $d$-electron occupations $n_d$
	computed at the experimental equilibrium volumes.}
\begin{tabular}{@{}rrrcrrcrrcrr@{}}\toprule
& \multicolumn{2}{c}{NiAs-type} & \phantom{abc}& \multicolumn{2}{c}{Rocksalt} & \phantom{abc} & \multicolumn{2}{c}{Wurzite} & \phantom{abc} & \multicolumn{2}{c}{Zincblende} \\
\cmidrule{2-3} \cmidrule{5-6} \cmidrule{8-9} \cmidrule{11-12} 
& LDA & LDA+RISB  && LDA & LDA+RISB  && LDA & LDA+RISB && LDA & LDA+RISB \\ \midrule
CrO & 4.31 & 4.27  && 4.35 & 4.25  && 4.2 & 4.14 && 4.2 & 4.14 \\
MnO & 5.33 & 5  && 5.25 & 5  && 5.08 & 5.06 && 5.16 & 5.13 \\
FeO & 6.38 & 6  && 6.34 & 6  && 6.2 & 6.17 && 6.22 & 6.2 \\
CoO & 7.43 & 7  && 7.42 & 7  && 7.24 & 7.21 && 7.27 & 7.22 \\
CoS & 7.45 & 7.45  && 7.48 & 7.49  && 7.52 & 7.55 && 7.52 & 7.54 \\
CoSe & 7.52 & 7.51  && 7.52 & 7.51  && 7.57 & 7.6 && 7.52 & 7.51 \\
\bottomrule
\end{tabular}
\label{n_d-table}
\end{table}

\begin{figure}[H]
	\begin{center}
		\includegraphics[width=0.99\linewidth]{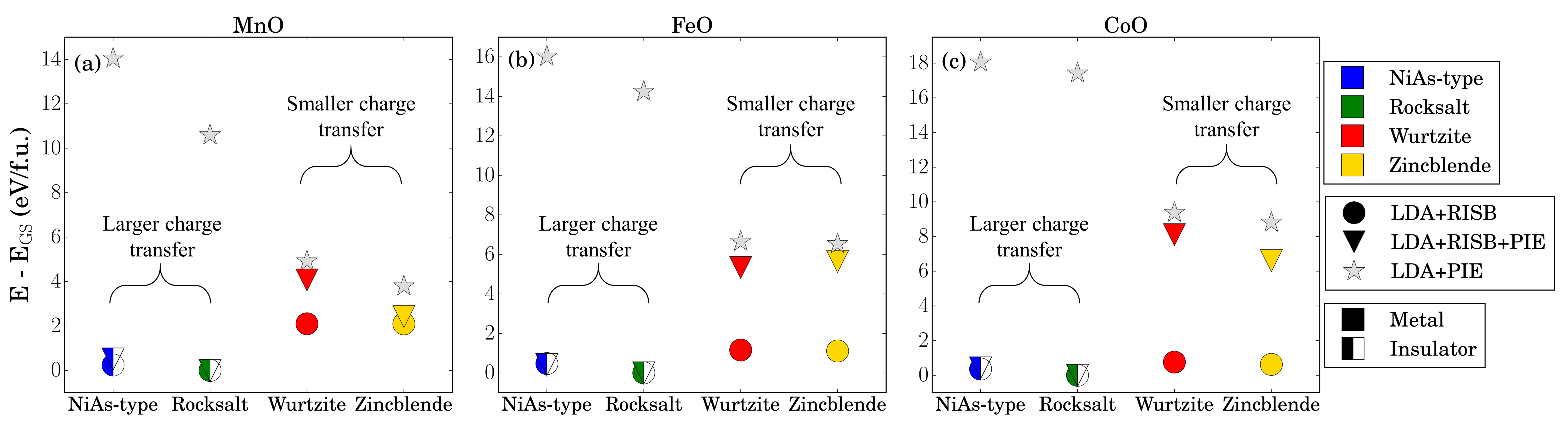}
		\caption{Energy differences between different crystal structures, 
		computed within
		LDA+RISB and the PIE model supplemented by the LDA+RISB $d$ occupations
		(LDA+RISB+PIE) and the LDA $d$ occupations (LDA+PIE).
		The results are shown for the Mott insulators MnO, FeO and CoO,
		and are computed at their respective experimental equilibrium volumes.
		The Rocksalt LDA+RISB+PIE and LDA+RISB energies are both
		conventionally assumed to be $0$ in all materials considered.
	}
		\label{electrostatics}
	\end{center}
\end{figure}

Interestingly, 
the values of $n_d$ are considerably influenced by 
the strong $d$-electron correlations, especially for the Mott 
insulating phases of MnO, FeO and CoO, which are realized in the NiAs-type and 
Rocksalt crystal structures.
For these materials,
we find that
the simple PIE electrostatic model supplemented by the
LDA+RISB occupations is sufficient to capture the correct LDA+RISB energy order
between the different crystal structures at their respective experimental volumes, while this is not 
the case if the LDA occupations are used, see Fig.~\ref{electrostatics}.
This analysis enables us to prove that the tendency of
electron correlations to favor energetically the Rocksalt and NiAs-type
structures is a consequence of Mott physics and electrostatics.
In fact, the key mechanism at play is that the metallic phases (which
are tetrahedrally coordinated) display
smaller charge transfer from the cation to the anion 
with respect to the Mott insulators
(which are octahedrally coordinated). 
Note that, not surprisingly,  
the PIE model is unable to capture the energy order 
of the metallic systems CrO, CoS and CoSe (not shown),
as the $d$ electrons retain a significant
mixed-valence character in these systems,
and the covalent effects are very important.

In summary, the main reason why the strong $d$-electron
correlations affect dramatically the crystal structure
of these materials is that
they influence substantially the charge transfer mechanism, which,
in turn, controls the electrostatic interactions.
This mechanism is particularly transparent in the Mott insulators
(MnO, FeO and CoO), where the role of the strong electron correlations
is magnified.
As we have shown, the energy order is considerably influenced by the
$d$-electron correlations also in the metallic systems examined.
However, the qualitative features of the chemical bonds in these metals
can not be captured by electrostatics alone,
as the covalent effects are also very important.
In this respect, it is also interesting to note
that for metallic systems (CrO, CoS and CoSe) the energy differences between 
crystal structures are much smaller than in the case of Mott insulators,
implying a more subtle balance between electrostatic and covalent 
effects --- which is accurately captured by a proper treatment of strong correlations.

\section*{Conclusions}

We have analyzed theoretically the ``choice'' of the ground state structure of 
6 transition-metal binary oxides and chalcogenides
among 4 different crystal structures,
utilizing LDA in combination with the
rotationally-invariant slave-boson (RISB) theory, 
finding very good quantitative agreement with the experiments.
These simulations demonstrated  that 
the subtle competing mechanisms underlying
the strongly-correlated regime 
influence dramatically the behaviour of the PES --- and,
in particular, the energy order between different crystal configurations ---
in all of the transition metal compound considered.

Remarkably, we found that the main qualitative features inherent in
the influence of the strong electron correlations on the crystal structure
can be understood in terms of a simple electrostatic model, 
based on the sole knowledge of the $d$-electron occupations.
This analysis enabled us to demonstrate 
that one of the key physical effects
which must be taken into account in order to predict
the correct energy ordering of these materials is the variation
of charge transfer induced by the 
strong $d$-electron correlations around the Mott point.
On the other hand, as expected,
electrostatics alone can not capture also the energy order 
of the metallic systems, where the $d$ electrons retain a significant
mixed-valence character, and the covalent effects are non-negligible.

These results indicate that simulation-based structure-prediction work of 
all transition-metal compounds (metals and insulors)
requires theoretical tools able to describe, at the same time: 
(i) the details of the bands structure, and
(ii) the $d$-electron correlations around the Mott transition
--- which influence
substantially the electronic structure and,
in particular, the charge-transfer mechanism in these materials.
For this purpose, taking into account the strong electron
correlations beyond single-particle mean-field
schemes, such as LDA and LSDA+U, is an absolute necessity.
In this respect, the LDA+RISB technique is particularly appealing,
as it is both sufficiently accurate and computationally convenient to be applied
to high-throughput computational materials design of this class of systems.

Another important conclusion arising from our study is that 
the typical energy differences between different
crystal structures, which have been all computed 
assuming paramagnetic solutions from the onset,
are generally much larger with respect energy scales characterizing magnetism 
in all of the materials considered (and most of the known transition metal oxides).
This observation indicates that 
magnetic order has generally a negligible effect on the total energy
with respect to the $d$-electron atomic scales originating 
the orbitally-selective correlations.
This physical insight results also in an appealing 
simplification from the computational standpoint,
as it suggests that it is possible 
to perform accurate structure-prediction work assuming from the onset
paramagnetic solutions (as we did in the LDA+RISB calculations
of the present work), i.e., without breaking translational invariance
or time reversal symmetry.

\section*{Methods}

The LDA and LDA+RISB calculations were performed utilizing the
DFT code WIEN2k~\cite{WIEN2k}.
The LDA+RISB solver was implemented
following Ref.~\cite{PRL_UO2_Lanata}.
The LAPW interface between WIEN2k and the RISB was implemented
as described in Ref.~\cite{Haule10}, utilizing
the fully-localized limit (FFL) double-counting functional~\cite{LDA+U}.
All calculation were performed setting 50000 $k$-points and $RKmax=8$.

\section*{Acknowledgements}

TH, VD, and NL were partially supported by the NSF grant DMR-1410132 and the National High Magnetic Field Laboratory. YY was supported by the U.S. Department of energy, Office of Science, Basic Energy Sciences, as a part of the Computational Materials Science Program. VS acknowledges support of the Center for the Next Generation of Materials by Design, an Energy Frontier Research Center funded by the U.S. Department of Energy, Office of Science, Basic Energy Sciences. The research was performed using computational resources sponsored by the Department of Energy's Office of Energy Efficiency and Renewable Energy and located at the National Renewable Energy Laboratory.
We thank Kevin John for providing us with Fig.~1.

\section*{Author contributions}

V.S. and V.D. initiated the project. All the authors contributed to the analysis and the interpretation of the results, and to writing the manuscript. N.L. and T.-H.L. equally contributed to this work. T.-H.L., N.L. and Y.-X.Y. performed the LDA+RISB calculations, while V.S. performed the DFT+U calculations.

\section*{Competing financial interests}
The authors declare no competing financial interests.


\pagebreak
\widetext
\begin{center}
\textbf{\large Supplemental Materials: Critical role of electronic correlations in determining crystal structure of transition metal compounds}
\end{center}
\section{LSDA+U analysis of $\text{CoS}$ and $\text{CoSe}$}

In Tables~\ref{CoS} and~\ref{CoSe} are shown the spin polarized GGA+U total energies and volumes for CoS and CoSe in four different structure types (NiAs-type, Rocksalt, Zincblenede, and Wurzite). Each total energy corresponds to the lowest energy spin configuration for the corresponding crystal structure at its respective relaxed volume. 
%
%
%
These results were obtained from an exhaustive enumeration of different spin configurations constructed on all symmetry inequivalent supercells with up
to four formula units (approximately 700 spin polarized GGA+U calculations for both CoS and CoSe).
The GGA+U calculations were performed following the approach 
described in Ref.~\cite{vladan_PRB:2012}, which employs PAW treatment of
valent electrons~\cite{bloechl_PRB:1994} 
and the Perdew-Burke-Ernzerhof (PBE) 
form for the exchange-correlation functional within the generalized gradient approximation to DFT~\cite{perdew_PRL:1996},
as implemented in the VASP code \cite{kresse_CMS:1996}.
Each total energy corresponds to the lowest energy spin configuration for that structure at the respective relaxed volume.



We observe that the
lowest-energy spin state predicts an incorrect energy ordering of the different crystal structures for all values of $U$.
For $U=0$, GGA+U predicts that
all solutions are metallic, and the NiAs-type
(which is thermodynamically stable experimentally)
has the highest energy.
For $U=6$ the GGA+U lowest-energy structure becomes Zincblende (remians wrong),
and, in addition, all solutions are insulators with volumes that are quite far from
the experimental one. 
The results do not improve for $U=12~eV$, as
the GGA+U lowest-energy structure becomes Wurtzite and also insulating.
We point out also that the energy differences between the GGA+U
lowest-energy structures and the NiAs-type structure
are considerably large for all values of $U$.
In particular, at $U=0$ we find about $200 meV/f.u.$
for CoS and about $300~meV/f.u.$ for CoSe.
For larger values of $U$, the energy difference between Rocksalt and NiAs-type decreases,
but the GGA+U energy minimum now becomes realized by the Zincblende and the
Wurzite stuctures. Actually, for all values of $U$ the energy preference is toward the tetrahedrally
coordinated structures. Even in case of $U=0$, for which the ground state is obtained by relaxing the initial rocksalt   
structure, the structural relaxations within the spin polarized GGA+U break the symmetry of the
rocksalt structure and transform it to the tetrahedrally coordinated structures with $Cm$ and $P4/mnm$ 
space groups for CoS and CoSe, respectively. For $U=6~eV$ and  $U=12~eV$ the rocksalt structure is stable. 
Therefore, the spin polarized GGA+U fails dramatically 
in reproducing known crystal structure and correct energy ordering in these systems.

These results demonstrate that
the strong electron correlations influence substantially the
behavior of both the electronic 
structure and the total energy of the correlated metals considered.

\begin{table}[H]\centering
	\ra{1.3}
	\caption{Spin polarized GGA+U total energies $E_{tot}$ and volumes $V$ for CoS in 4 different structure types (NiAs-type, Rocksalt, Zincblenede, and Wurzite), computed at 3 different values
	of the Hubbard $U$.
	Each total energy corresponds to the lowest energy spin configuration for the corresponding crystal structure at its respective relaxed volume. The total energy is always referenced to the ground state structure for a given $U$ value. }
	\begin{tabular}{@{}rrrcrrcrrcrr@{}}\toprule
		& \multicolumn{2}{c}{$U=0\,eV$} & \phantom{abc}& \multicolumn{2}{c}{$U=6\,eV$} & \phantom{abc} & \multicolumn{2}{c}{$U=12\,eV$} 
		\\
		\cmidrule{2-3} \cmidrule{5-6} \cmidrule{8-9} 
		& $E_{tot}\,[eV/f.u.]$ & $V\,[eV/f.u.]$  && $E_{tot}\,[eV/f.u.]$ & $V\,[eV/f.u.]$  && $E_{tot}\,[eV/f.u.]$ & $V\,[eV/f.u.]$ \\ 
		NiAs-type & 0.164 & 24.92  && 0.216 & 32.16  && 0.179 & 33.09 \\ 
		Rocksalt & 0.000 & 28.92  && 0.235 & 32.15  && 0.184 & 33.06 \\ 
		Wurzite & 0.050 & 30.56  && 0.000 & 40.59  && 0.002 & 42.23 \\ 
		Zincblende & 0.003 & 28.59  && 0.019 & 40.88  && 0.000 & 42.34 \\ 
		\bottomrule
	\end{tabular}
	\label{CoS}
\end{table}

\begin{table}[H]\centering
	\ra{1.3}
	\caption{Spin polarized GGA+U total energies $E_{tot}$ and volumes $V$ for CoSe in 4 different structure types (NiAs-type, Rocksalt, Zincblenede, and Wurzite), computed at 3 different values
	of the Hubbard $U$.
	Each total energy corresponds to the lowest energy spin configuration for the corresponding crystal structure at its respective relaxed volume. The total energy is always referenced to the ground state structure for a given $U$ value. }
	\begin{tabular}{@{}rrrcrrcrrcrr@{}}\toprule
		& \multicolumn{2}{c}{$U=0\,eV$} & \phantom{abc}& \multicolumn{2}{c}{$U=6\,eV$} & \phantom{abc} & \multicolumn{2}{c}{$U=12\,eV$} 
		\\
		\cmidrule{2-3} \cmidrule{5-6} \cmidrule{8-9} 
		& $E_{tot}\,[eV/f.u.]$ & $V\,[eV/f.u.]$  && $E_{tot}\,[eV/f.u.]$ & $V\,[eV/f.u.]$  && $E_{tot}\,[eV/f.u.]$ & $V\,[eV/f.u.]$ \\ 
		NiAs-type   & 0.283 & 29.36  && 0.189 & 37.34  && 0.198 & 38.68 \\ 
		Rocksalt     & 0.000 & 39.05  && 0.232 & 37.32  && 0.220 & 38.77  \\ 
		Wurzite       & 0.376 & 36.56  && 0.000 & 47.58  && 0.000  & 49.42 \\ 
		Zincblende & 0.311 & 33.36  && 0.016 & 47.84  && 0.001 &  49.54  \\ 
		\bottomrule
	\end{tabular}
	\label{CoSe}
\end{table}

\section{LDA+RISB analysis of energy order for all structures.}

\begin{figure}[H]
	\begin{center}
		\includegraphics[width=0.95\linewidth]{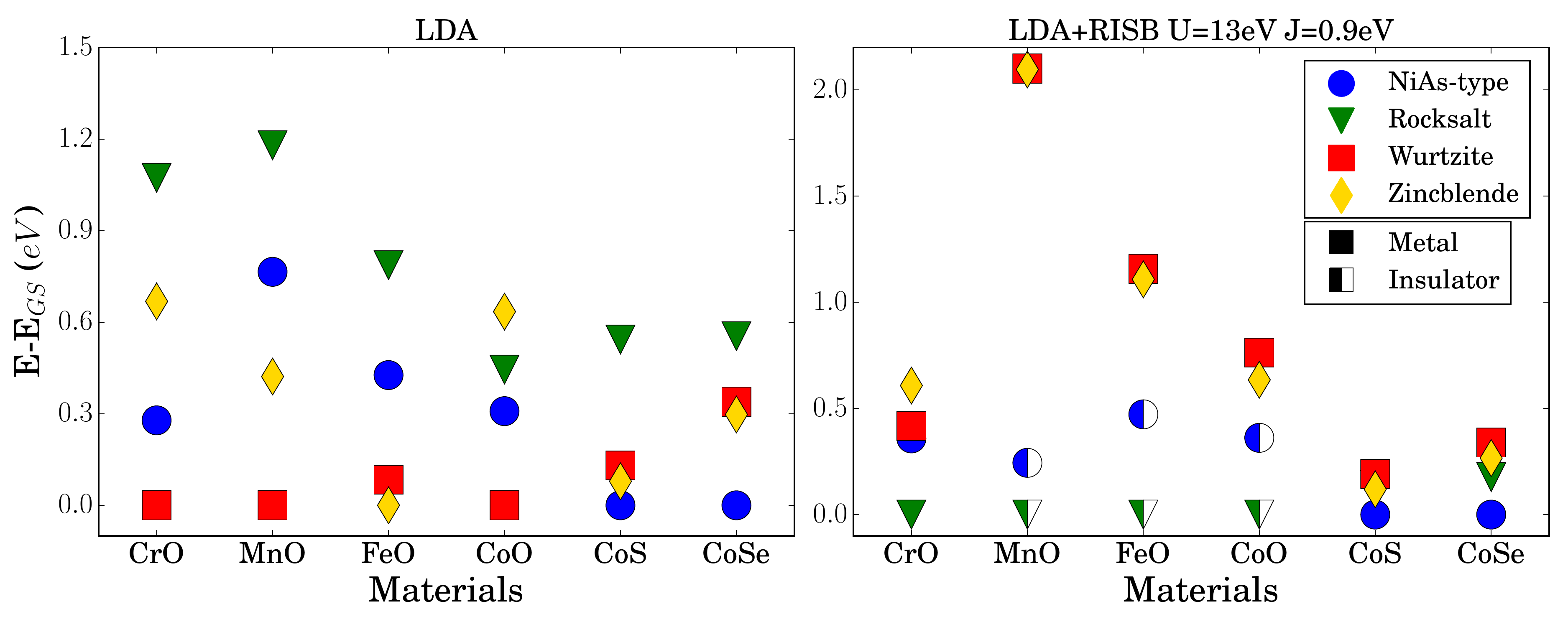}
		\caption{Total energy differences
			of CrO, MnO, FeO, CoO, CoS and CoSe in all of
			the crystal structures considered at the respective equilibrium volumes at ambient conditions.
			%
			The blue circle, green triangle, red square, and yellow diamond curves correspond to NiAs-type, rocksalt, wurtzite, and zincblende structures, respectively.
			The metallic and insulating solution are labeled by filled and half-filled symbols, respectively.
			}
		\label{FigS3}
	\end{center}
\end{figure}

In Fig.~\ref{FigS3} are shown the LDA+RISB total energy differences
of CrO, MnO, FeO, CoO, CoS and CoSe computed for all of
the crystal structures considered at the experimental
equilibrium volumes realized at ambient conditions.
The lowest energy of each material is conventionally considered $0$.
The metallic and insulating solution are labeled by filled and half-filled symbols, respectively.

As discussed in the main text,
MnO, FeO and CoO are Mott insulators in the octahedrally coordinated structures (Rocksalt and NiAs-type), 
while they are metals in the tetrahedrally coordinated structures
(Wurzite and Zincblende).

We observe that the RISB correction modifies dramatically 
the LDA energy order of all systems.
For the metallic materials (CrO, CoS and CoSe)
the energy differences between 
crystal structures are much smaller than in the Mott insulators,
implying a more subtle balance between electrostatic and covalent 
effects.

\end{document}